\documentclass[amsmath,twocolumn,showpacs,aps,prl]{revtex4-1}
\usepackage{bm}
\usepackage{graphics}
\usepackage{graphicx}

\begin{document}

\title{Magnon mechanism of Josephson coupling in SFS structures}
\author{A. Yu. Zyuzin}

\affiliation{A.F. Ioffe Physico-Technical Institute of
Russian Academy of Sciences, 194021 St. Petersburg, Russia}

\pacs{74.45.+c, 75.40.Gb}

\begin{abstract}
It is shown that Josephson coupling in SFS junction due to electron-magnon interaction remains at a distance, when the usual proximity effect decreases exponentially. We obtain expression for the Josephson energy, which contain the parameters of the magnon spectrum and allow to estimate the value of the maximum superconducting current.
\end{abstract}
\maketitle

The expected physics of the proximity effect in the structures of S-superconductor-ferromagnetic metal \cite{bib:Buzdin1} is based on the fact that the wave function of a Cooper pair in a metal oscillate with the distance from the boundary with the superconductor. This is the analogue of the behavior in the superconducting LOFF state \cite{bib:Fulde, bib:Larkin}. The presence of impurity scattering in the metal leads to an exponential decrease of the wave function of Cooper pair on the oscillation length, equal to $L_{h}=\sqrt{D/h}$, where $D$ is the diffusion coefficient, and $h$ is the energy of the ferromagnetic splitting. Experimental investigation of this pattern is well established \cite{bib:Ryazanov}, see also the review \cite{bib:Buzdin2,bib:Izum,bib:Bergeret}.

Recent experiments on SF structures found \cite{bib:CrO2, bib:Ho,bib:Norman}, that the Cooper pairs in a superconductor-magnetic metal penetrate to a distance much greater than the length $L_ {h}$. The explanation is that in these structures magnetic state is characterized by a noncollinear ordering. In \cite{bib:Ho} it is a helical ordering, in \cite{bib:Norman} it is artificially created in magnetic multilayers. In this case, odd frequency triplet state penertates into magnetic metal \cite{bib:Bergeret, bib:Volkov}. The spatial symmetry of this state determines its insensitivity to a potential scattering.

Mesoscopic fluctuations of Josephson current are preserved on all scales associated with both the elastic interactions and ferromagnetic splitting. Therefore, in SFS junction, in which distribution of the of the superconducting phase difference is correlated with the mesoscopic fluctuations, there is an average Josephson current \cite{bib:Zyuz} even at thicknesses $d$ much larger than $L_{h}$.

In this paper we study the influence of electron-magnon interaction on the Josephson energy of the SFS junction with thickness $d>>L_{h}$. Usually, when considering the problem of the SFS contact, ferromagnet is modeled as a metal with a built-in spin splitting. This approximation  neglects the fact that due to electron-magnon interaction electronic states with a given spin projection are not eigenstates. Therefore, the wave function of Cooper pairs penetrating into the ferromagnet from S-superconductor contains a component, which does not oscillate in the exchange field and, as it is shown by calculation, does not decreases exponentially over a length $L_{h}$. Here we calculate corrections to the thermodynamic potential of SFS junction due to electron-magnon interaction.

We consider the s-d model for a ferromagnet with Hamiltonian
\begin{equation}\label{sd}
H_{sd}=J\int d{\bf r}\psi_{i}^{+}({\bf r})\pmb\sigma_{i,j}\psi_{j}({\bf r}){\bf S}({\bf r})
\end{equation}
Here $\pmb\sigma$ are the Pauli matrices,
${\bf S}({\bf r})=\sum_{k}\delta({\bf r}-{\bf r}_{i}){\bf S}_{k}$
is the density operator of spins, localized at the points ${\bf r}_{i}$.

We assume ferromagnetic ordering of localized spins in the direction $z$ and corresponding splitting of the electron spectrum of ferromagnetic $h =J\langle S\rangle$. $\langle S\rangle$ is the density of localized spins.

The transverse part of the Hamiltonian (\ref{sd}) with the help of Nambu operators
\begin{equation}
 \Psi_{1}(\textbf{r})=
\begin{pmatrix}
\psi_{\uparrow}({\bf r})\\
\psi_{\downarrow}^{+}({\bf r})            
\end{pmatrix},
\Psi_{2}({\bf r})=
\begin{pmatrix}
\psi_{\downarrow}({\bf r}) \\
\psi_{\uparrow}^{+}({\bf r})             
\end{pmatrix},
\end{equation}
might be written as
\begin{equation}
H_{\perp}=\frac{J}{2}\sum_{i,j=1,2}\int d{\bf r}\Psi^{+}_{i}({\bf r})\left(\sigma_{i,j}^{x}S^{x}({\bf r})+\sigma_{i,j}^{y}S^{y}({\bf r})\right)\sigma^{z}\Psi_{k}({\bf r})
\end{equation}
The second-order perturbation correction to the thermodynamic potential has the form
\begin{widetext}
\begin{equation}\label{TP}
\delta\varOmega=-\frac{J^2}{8}{\int_{0}}^{1/T}d\tau_{1}d\tau_{2}\int d{\bf r}_{1}d{\bf r}_{2}\sum_{\alpha,\beta=x,y}D^{\alpha,\beta}(\tau_{12},{\bf r}_{1},{\bf r}_{2}) 
\sigma_{i,j}^{\alpha}\sigma_{j,i}^{\beta}\langle\sigma^{z}G_{i}(\tau_{21},{\bf r}_{2},{\bf r}_{1})\sigma^{z}G_{j}(\tau_{12},{\bf r}_{1},{\bf r}_{2})\rangle 
\end{equation} 
\end{widetext}

The integration is over the ferromagnetic contact region $\rvert x\lvert<d/2 $. We assume that it is in the $y,z$ plane.
Here $D_{i,k}(\tau_{12},{\bf r}_{1},{\bf r}_{2})$ is the Matsubara Green's function for the spin operators
\begin{equation}
D_{i,k}(\tau_{12},{\bf r}_{1},{\bf r}_{2})=-\{T_{\tau} S^{i}(\tau_{1},{\bf r}_{1})S^{k}(\tau_{2},{\bf r}_{2})\}
\end{equation} 

Electron Green's function, defined as
\begin{equation}
G_{i}(\tau_{12},{\bf r}_{1},{\bf r}_{2})=-\{T_{\tau}\Psi_{i}(\tau_{1},{\bf r}_{1})\Psi^{+}_{i}(\tau_{2},{\bf r}_{2})\} 
\end{equation}
is the $2\times 2$ matrix.

Operators $\Psi_{1}({\bf r})$ and $\Psi_{2}({\bf r})$ differ only in the direction of the electron spins, therefore
$G_{1}(\tau,{\bf r}_{1},{\bf r}_{2},h)=G_{2}(\tau,{\bf r}_{1},{\bf r}_{2},-h)$

The figure shows a diagram corresponding to the thermodynamic potential (\ref{TP}), averaged over the random potential. We apply the standard technique of averaging described in the review \cite{bib:Buzdin2, bib:Izum, bib:Bergeret} and the references cited therein.

The averaging is performed over the scattering potential, which in the Nambu representation is $\sigma^{z}V({\bf r})$. It is assumed that $\langle V({\bf r})V({\bf r}')\rangle=\frac{1}{2\pi \nu_{0}\tau}\delta({\bf r}-{\bf r}')$, where $\tau$ and $\nu_{0}$ are the mean free path and density of states per spin at the Fermi level.

\begin{figure}
\includegraphics[width=0.8\linewidth]{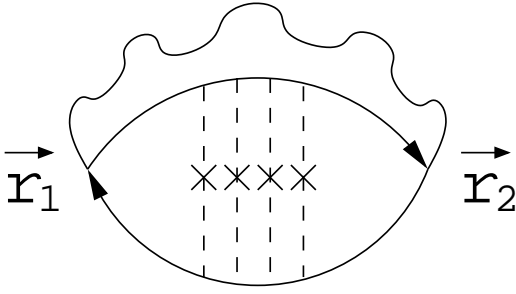}
\caption{The diagram corresponding to magnon contribution to the thermodynamic potential, averaged over the random scattering potential. The wavy line in the figure denotes the magnon propagator. The solid lines correspond to $2\times 2 $ electron Green's function, averaged over the random potential. Dotted lines correspond to the impurity scattering. It is shown element of the diffusion propagator. \label{fig}}
\end{figure}

The equation for the electron Green's function, averaged over the scattering potential has the form
\begin{equation}\label{FG}
(i\epsilon_{n}-h-H_{0}\sigma^{z}+\frac{i}{2\tau}\hat{g}_{1}({\bf r}))\langle G_{1}({\bf r},{\bf r}',\epsilon_{n})\rangle=\delta({\bf r}-{\bf r}')
\end{equation}
Here, $\epsilon_{n} =(2n +1)\pi T$ is the Matsubara frequency, $T$ is the temperature, and $H_{0}$ is the Hamiltonian of free electrons.

The off-diagonal elements of the Green's function describe the penetration of Cooper pairs in the normal metal over the length of the $L_{h}=\sqrt{D/h}$. In our case, the thickness of the ferromagnetic of metal $d>>L_{h}$, therfore both superconductor-ferromagnet interfaces might be a  considered independently.

Near the boundary with the superconductor,which order parameter has a phase $\varphi$, the matrix $\hat{g}_{1}$ can be written as

\begin{equation}\label{uzadel}
\hat{g}_{1}({\bf r})\equiv\frac{i\sigma^{z}}{\pi\nu_{0}}\langle G_{1}({\bf r},{\bf r},\epsilon_{n})\rangle=
\begin{pmatrix}
g_{1}&-if_{1}e^{i\varphi}\\
if_{1}e^{-i\varphi}&-g_{1}                
\end{pmatrix}
\end{equation} 
The matrix elements are related by $g_{1}^2 +f_{1}^2 =1$.

At a distance from the boundary larger than the $L_{h}$, or in the case of a weak proximity effect, the matrix elements are $\rvert f_{1}\lvert<<1$, $g_{1}=sign(\epsilon_{n})$.

The function $f_{1}({\bf r},\epsilon_{n})$ at $\rvert\epsilon_{n}\tau\lvert<<1$ и $h\tau<1$ might be determined from the Usadel equation, which in the case of a weak proximity effect has form
\begin{equation}
 (-D\nabla^2+2(\epsilon_{n}+ih)sign(\epsilon_{n}))f_{1}({\bf r},\epsilon_{n})=0
\end{equation} 
Here $D$ is the electron diffusion coefficient.
Note that $f_{2}({\bf r},\epsilon_{n}) $ might be obtained from $f_{1}({\bf r},\epsilon_{n})$ by changing sign of $h\rightarrow -h$.

In the case of low transparency of the SF interface,  the boundary conditions at the contact between the superconductor-ferromagnetic metal is \cite{bib:kuprian}
\begin{equation}
\gamma {\bf n}\nabla f_{1}({\bf r},\epsilon_{n})=\frac{\Delta}{\sqrt{\Delta^2+\epsilon_{n}^2}}
\end{equation} 
Here $(\gamma)^{-1}$ is the ratio of the resistivity of the ferromagnet to the resistance per unit area of ​​surface. $\bf n$ is the normal to the surface. $\Delta$ is the modulus of the superconducting order parameter.

In a bulk of ferromagnetic metal the diffusion propagators appearing in susceptibility (see figure)
\begin{equation}\label{Pi}
\varPi\equiv\langle Sp(\sigma_{z}G_{1}({\bf r}_{1},{\bf r}_{2},\epsilon_{n})\sigma_{z}G_{2}({\bf r}_{2},{\bf r}_{1},\epsilon_{n'}))\rangle,
\end{equation} 
which are proportional to $(1\pm\sigma^{(1)}_{z})(1\mp\sigma^{(2)}_{z})$ do not contain $h$. 
Accordingly, only those contributions are not damped at the length $L_{h}$ and should be considered for the calculation of $\delta\varOmega$ for junction with $d>>L_{h}$. In this case the frequencies $\epsilon_{n}$ and $\epsilon_{n'}$ must satisfy the condition $\epsilon_{n}\epsilon_{n'}>0$.

$\hat{g}_{i}({\bf r})$ is slow varying function of coordinates at the mean free path distance.
Neglecting it's gradients we obtain for vertex part of (\ref{Pi}) expression
\begin{equation}\label{vertex}
\int \frac{d^3{\bf p}}{(2\pi)^3}\langle G_{1}({\bf r},{\bf p},\epsilon_{n})\rangle
\sigma^{z}\langle G_{2}({\bf r},{\bf p},\epsilon_{n'})\rangle=\pi \nu_{0}\tau\sigma^{z}(1-\hat{g}_{1}\hat{g}_{2})
\end{equation} 

Substituting in (\ref{vertex}) definition (\ref{uzadel}) in the case of a weak proximity effect we obtain for $\epsilon_{n}\epsilon_{n '}>0$
\begin{equation}\label{vertex1}
 \sigma^{z}(1-\hat{g}_{1}({\bf r})\hat{g}_{2}({\bf r}))\simeq -i
\begin{pmatrix}
0&e^{i\varphi}\\
e^{-i\varphi}&0            
\end{pmatrix}
(f_{1}({\bf r},\epsilon_{n})-f_{2}({\bf r},\epsilon_{n'}))
\end{equation} 

Thus, the entering into long-range contribution to the $\varPi$ vertex parts are different from zero only near the surfaces of contact.

The equation for the diffusion propagator, neglecting Andreev reflection ($ \hat{g}_{1}({\bf r})=sign(\epsilon_{n})\sigma^z $) is obtained using the Green function (\ref{FG}) with $\epsilon_{n}\epsilon_{n'}>0$. It has the form $\frac{\sigma_{z}^{(1)}\sigma_{z}^{(2)}-1}{4\pi\nu_{0}\tau^2}P({\bf r},{\bf r}',\Omega_{n,n'})$, where
\begin{equation}\label{dif}
(-D\nabla^2+\rvert \Omega_{n,n'}\lvert)P({\bf r},{\bf r}',\Omega_{n,n'})=\delta({\bf r}-{\bf r}')
\end{equation} 
Here, $\Omega_{n,n'}\equiv \epsilon_{n}+\epsilon_{n'}$.

In the limit $d>>L_{h}=\sqrt{D/h}$ in the integration over the spatial coordinates in (\ref{TP}) slowly varying function of the coordinates $D$ and $P$ can be set equal to their values ​​in the $x=\pm d/2$. The integration of vertex parts over $x$ is reduced to the replacement
 \begin{align}
(f_{1}({\bf r}_{1},\epsilon_{n})-f_{2}({\bf r}_{1},\epsilon_{n'}))(f_{2}({\bf r}_{2},\epsilon_{n'})-f_{1}({\bf r}_{2},\epsilon_{n}))\rightarrow \nonumber\\
\left(\frac{D}{2h\gamma}\left(\frac{\Delta}{\sqrt{\Delta^2+\epsilon_{n}^2}}+\frac{\Delta}{\sqrt{ \Delta^2+\epsilon_{n'}^2}}\right)\right)^2 
\end{align}

There are two contributions to the thermodynamic potential.

If ${\bf r}_{1}$ and ${\bf r}_{2}$ are on opposite surfaces of junction, than the phase dependent susceptibility has the form

\begin{align}\label{Pi1}
\delta_{1}\varPi({\bf r}_{1},{\bf r}_{2})&=\pi\nu_{0}\left(\frac{D}{2h\gamma}\left(\frac{\Delta}{\sqrt{\Delta^2+\epsilon_{n}^2}}+\frac{\Delta}{\sqrt{ \Delta^2+\epsilon_{n'}^2}}\right)\right)^2\nonumber\\
\times& P({\bf r}_{1},{\bf r}_{2},\Omega_{n,n'})\cos\varphi_{12}  
\end{align} 

When the coordinates ${\bf r}_{1}$ and ${\bf r}_{2}$ belong to the same surface, phase dependent contribution arises after taking into account Andreev reflection from the opposite surface. Graphically, this means inserting a Hikami block containing anomalous part of the Green's functions in the diffusion propagator shown in Fig.
Rate of the reflection is  $D\nabla f_{1}({\bf r})\nabla f_{2}({\bf r})$. As a result,

\begin{widetext}
\begin{equation}\label{Pi2}
\delta_{2}\varPi({\bf r}_{1},{\bf r}_{2})=-\frac{\pi D\nu_{0}}{2}\left(\frac{D}{2h\gamma}\left(\frac{\Delta}{\sqrt{\Delta^2+\epsilon_{n}^2}}+\frac{\Delta}{\sqrt{ \Delta^2+\epsilon_{n'}^2}}\right)\right)^2
\cos2\varphi_{12}\int d{\bf r} P({\bf r}_{1},{\bf r},\Omega_{n,n'})\nabla f_{1}({\bf r})\nabla f_{2}({\bf r}) P({\bf r},{\bf r}_{2},\Omega_{n,n'})
\end{equation}
\end{widetext}
The integration over ${\bf r}$ is near surface, which is opposite to that of the ${\bf r}_{1}$ and ${\bf r}_{2}$. Here again we can put the diffusion propagators are equal to their values ​​at the surface and integrate $D\nabla f_{1}({\bf r})\nabla f_{2}({\bf r})$ over $x$.

Since when $h>>\rvert\epsilon_{n}\lvert,\rvert\epsilon_{n'}\lvert$ $\delta_{1}\varPi$ and $\delta_{2}\varPi$ are even functions $h$, the thermodynamic potential depends on a combination of spin Green's functions as
\begin{equation}
 D_{x,x}(q,\epsilon_{n}-\epsilon_{n'})+D_{y,y}(q,\epsilon_{n}-\epsilon_{n'})=\frac{2\langle S\rangle E(q)}{E(q)^2+(\epsilon_{n}-\epsilon_{n'})^2}
\end{equation} 
Here $E(q)=E_a+D_{s}q^2$ is the magnon energy, $E_a$ takes into account the anisotropy energy and the external magnetic field, $\langle S\rangle$ is the spin density.

Let consider the limits of 1). weak anisotropy $\sqrt{D_{s}/E_{a}}>L_{h}$, when magnon propagator is slowly varying on the length $L_{h}$, and 2). temperature, such that $T>D_{s}/d^2, E_{a}$. 

In this case summation over frequencies in thermodynamic potential (\ref{TP}) might be restricted by terms with $\epsilon_{n}=\epsilon_{n'}$
  
Normally $D>>D_{s}$, so in this limit there might be any possible relationship between the thickness $d$ and the coherence length $\sqrt{D/T}$.

After substituting in (\ref{TP}) expressions (\ref{Pi1}), (\ref{Pi2}) and the magnon propagator, calculated for an infinite medium, and integrating over the SF surfaces, we finally obtain
\begin{equation}
\delta\varOmega=\delta_{1}\varOmega+\delta_{2}\varOmega, 
\end{equation} 
where per unit area
\begin{align}\label{omega1}
\frac{\delta_{1}\varOmega}{S}=\frac{D\nu_{0}\cos\varphi_{12}}{8\gamma^{2}D_{s}\langle S\rangle}T^2\sum_{\epsilon_{n}}\frac{\Delta^2}{ \Delta^2+\epsilon_{n}^2}\nonumber\\
\times\int_{1}^{\infty} \frac{dt}{t}
\exp\left(-t\left(\sqrt{\frac{\rvert 2\epsilon_{n}\lvert }{D}}+\sqrt{\frac{E_a}{D_{s}}}\right)d\right)
\end{align}
and
\begin{align}\label{omega2}
\frac{\delta_{2}\varOmega}{S}=-\frac{D\nu_{0}\cos 2\varphi_{12}}{64\gamma^{4}D_{s}\langle S\rangle}\sqrt{\frac{DD_{s}}{hE_{a}}}T^2\sum_{\epsilon_{n}}\frac{\Delta^4}{ (\Delta^2+\epsilon_{n}^2)^2}\nonumber\\
\times\int_{1}^{\infty} \frac{dt}{t}
\exp\left(-2t\sqrt{\frac{\rvert 2\epsilon_{n}\lvert }{D}}d\right)
\end{align}
In deriving these expressions we used the relation $h =J\langle S\rangle$ between the sd interaction constant and the value of ferromagnetic splitting. Integration of diffusion and magnon propagators over surfaces is reduced to integration over $t$.

Note that minimum of $\frac{\delta_{1}\varOmega}{S}$ corresponds to $\varphi=\pi$ state, and minimum $\frac{\delta_{2}\varOmega}{S}$ is achieved at $\varphi=0,\pi$.

The spin-orbit interaction has two consequences. The presence of a gap in the magnon spectrum, corresponding to the anisotropy energy $E_{a}$ for $d>\sqrt{D_ {s}/E_ {a}}$ leads to the factor $\exp(-d\sqrt{E_{a}/D_{s}})$ in the expression (\ref{omega1}). The spin-orbit scattering of conduction electrons can be accounted for by the substitution $\rvert 2\epsilon_{n}\lvert\rightarrow\rvert 2\epsilon_{n}\lvert+\tau^{-1}_{so}$ in the exponents in (\ref{omega1}) and (\ref{omega2}). $\tau_{so}$ is the spin relaxation time of conduction electrons due to spin-orbit scattering. When $d$ is greater than the length of the spin relaxation of conduction electrons $L_{so}=\sqrt{D\tau_{so}}$ the first contribution decreases as $\exp(-d/L_{so})$. In this case, the second contribution decreases with increasing spin-orbit scattering faster than the first as $\exp(-2d/L_{so})$.

Let estimate the value of the maximum superconducting current corresponding to the expressions (\ref{omega1}) and (\ref{omega2}) with $\sqrt{D/T}, L_{so}>d $, $\Delta\gtrsim T$, when the sum over the frequencies gives a contribution of order of unity.

At specific resistance of the ferromagnet $\sim 0.1 \mu$ом$\times$cm, the factor $D\nu_{0}$ is $\sim 2\times 10^{10}cm^{-1}$. For 3d metals, magnon spectra have $D_{s}\sim 10^{-17}eV$ cm$^2$ \cite{bib:3d}. At spin density $\langle S\rangle\sim 10^{22}$ cm$^{-3}$ and temperature $\sim 1^{0}K$ we have $\frac{D\nu_{0}}{8D_{s}\langle S\rangle}T^2\sim 2^{0}K$.
$\gamma$ is the ratio of the mean free path in the ferromagnet to the boundary transmission coefficient \cite{bib:kuprian}. If $\gamma\sim 10^{-5}$cm than for area $S\sim 10^{-8}$cm$^2$ maximum superconducting current is a few $\mu$A.

Note that because of the smallness of the $D_{s}$ the length $\sqrt{D_{s}/E_a}$ may be small even for weak anisotropy, or magnetic field. In this case, contribution to (\ref{omega1}) is less than the contribution of (\ref{omega2}).

This work was supported by RFFI under Grant No.
12-02-00300-A.

\end{document}